# Smart-optimism. Uncovering the Resilience of Romanian City Halls in Online Service Delivery

**Catalin Vrabie**
**Faculty of Public Administration**
**National University of Political Studies and Public Administration**
**Bucharest, Romania**
catalin.vrabie@snspa.ro

**Abstract**

Recent technological advancements have significantly impacted the public sector's service delivery. Romanian city halls are embracing digitalization as part of their development strategies, aiming to deploy web-based platforms for public services, enhancing efficiency and accessibility for citizens. The COVID-19 pandemic has expedited this digital shift, prompting public institutions to transition from in-person to online services. This study assesses the adaptability of Romanian city halls to digitalization, offering fresh insights into public institutions' resilience amidst technological shifts. It evaluates the service provision through the official web portals of Romania's 103 municipalities, using 23 indicators for measuring e-service dissemination within local contexts. The research reveals notable progress in the digital transformation of services over time (2014-2023), with a majority of municipalities offering online functionalities, such as property tax payments, public transportation information, and civil status documentation. It also discovers disparities in service quality and availability, suggesting a need for uniform digitalization standards. The findings enlighten policymakers, assist public institutions in advancing digital service delivery, and contribute to research on technology in public sector reform.

## 1. Introduction

The COVID-19 pandemic has brought about significant changes in various aspects of our lives, including the way cities operate and the services they provide to their citizens. The pandemic has highlighted the importance of resilient and adaptive public institutions, especially in the context of smart cities. A smart city is a concept that emphasizes the use of technology and data to improve the quality of life for citizens while ensuring sustainable development (Anthony & Andreas, 2023; British Standards Institution, 2014; Cisco, 2012).

The pandemic has had a significant impact on the use of technology and its role in the resilience of public institutions in smart cities and prompted a strong push towards the digitalization of services, replacing on-site activities of public institutions (Matthias, et al., 2020; Kuhlmann, et al., 2021; MÜLLER-TÖRÖK & PROSSER, 2021). The need for physical distancing and remote communication during the pandemic has highlighted the importance of technology in maintaining the continuity of public services in times of crisis. In this context, smart cities have the potential to play a critical role in the delivery of public services, especially in times of crisis. The resilience of public institutions to adapt to technological change and new circumstances is crucial for the success of smart city initiatives (Anthony & Andreas, 2023; SCHACHTNER, 2021; Šiugždinienė, et al., 2017; Nature, 2022; WHO, 2023).

In this article, our objective is to explore the robustness of public institutions within the framework of smart cities following the COVID-19 pandemic. We concentrate on the adaptability of public institutions to technological advancements and their capacity to offer services through remote channels (online and mobile). Our goal is to present new evidence highlighting the endurance of public institutions amid technological shifts and the contribution of smart web platforms in this regard. By assessing the degree of digitalization in public services across Romanian municipalities (as of 2023) and comparing it to the progress of web initiatives from 2014 and 2019 respectively, we aspire to add to the ongoing discussion about the future of public services in the digital era, the function of public institutions in city transformations, and the steadfastness of public institutions during crises.

Following the introduction, a literature review section will be presented, highlighting key scientific approaches and research studies that discuss the concept of smart cities and their relationship with online and mobile technologies. This will help situate the current study within the existing body of literature. The article will then proceed with Digitalization framework in Romania, which introduces the digitalization framework in Romania, providing context for the research. The research methodology, will offer details on data collection methods and the tools employed, leading to results. This section is split into three subsections, each addressing distinct results of the research. Study limitations is followed by the findings. Lastly, the discussion and conclusion section will convey the author's perspective regarding the study's findings and implications.

## 2. Literature review

This section outlines the evolution of e-government, beginning with basic information dissemination on websites and progressing to interactive platforms for citizen communication and electronic document management (Vrabie, 2016). Subsequently, the next phase saw the integration of online payments and active citizen participation through social media and virtual platforms. The third phase, characterized by the adoption of AI in administrative tasks (Vrabie & Dumitrascu, 2018; Klievink & Janssen, 2019; TIMAN, et al., 2021) emphasizes the role of advanced technologies in improving decision-making and strengthening citizen-public institution relationships (Vrabie, 2022).

As early as the year 2000, the theoretical approaches of (Orlikowski, 2000) and (Czarniawska & Guje, 2005) are based upon the idea of environmental pressures as being the reasons for building Web platforms and implementing electronic services. Focusing on the concepts of "organization" and "structure", (Wanda Orlikowski & Stephen Barley, 2001) hope to bring light to the process of technology development and organizational change, thus succeeding to better stimulate the implementation of electronic services.

A good model of studying digitalization at local level is provided by Marc Holzer who completed a series of analysis over municipalities worldwide during 2005 and 2016 namely „Digital Governance in Municipalities Worldwide" (Holzer & Aroon P. Manoharan, 2016) while Journal of Web Semantics by Elsevier is providing a series of articles (issues 2016–2023) very much connected with the content and design of Web sites.

Upon examining the research presented in Government Information Quarterly (issues 2020-2023), and International Journal of Web Services Research (issues 2020-2023), it can be inferred that the majority of attention is directed towards e-government in a general sense, with limited emphasis on its implementation on a large scale, such as at a country level, with a detailed focus as our study entails.

The idea of integrating Web services was promoted by (Tapscott, 2008) in his memorable book "Grown Up Digital: How the Net Generation is Changing Your World" and by (Homburg & Andres Dijkshoorn, 2011) who wrote about "Diffusion of Personalized E-Government Services among Dutch Municipalities". This type of services envisages the citizens' interaction with the public administration, and through the procedures of authentication and assignment to a profile, the interaction between public services' provider and the user becomes the *one-to-one* type.

According to the literature on public management, the utilization of web applications can be an effective solution for tackling these challenges and promoting the establishment and maintenance of good governance (Klievink & Janssen, 2019; Deng, et al., 2019; Cai & Zhu, 2021; Alves & Gonçalves, 2021; Nesti, 2018). As an illustration, the use of Web applications can enhance the quality and efficiency of public service delivery (Maiti, et al., 2021; THAKHATHI & D, 2022), automating administrative tasks (de Lange-Ros, et al., 2018), and supporting decision-making processes (Shafique, et al., 2020; SCHACHTNER, 2021) especially when empowered by AI applications. Moreover, online platforms can enhance transparency and accountability, as well as increase citizen participation and engagement (Etscheid, 2019; Kolkman, 2020).

In 2021, (Piaggesi, 2021) conducted research on the future of connectivity and presented an overview of Latin America and recommended that for a smooth transition to e-government 3.0, it is crucial for the government to play a significant role in providing universal service. (Verma, 2022) conducted a thorough bibliometric review of 353 research articles published between 2010 and 2021 to assess the effectiveness of public servants. The study concluded that the use of smart technologies is contributing to the development of smarter governance structures across society.

Furthermore, in a systematic literature review (Madan & Ashok, 2023) identified contextual variables as key factors that impact the adoption of online services, as discussed in the literature. The study concluded that governance maturity plays a vital role in effectively managing IT implementation. Additionally (Ahn & Yu-Che Chen, 2022) investigated the perceptions of public employees concerning the utilization of technology in government operations. The authors discovered that government employees maintain a favorable attitude towards the advantages and potential of technology in the public sector. They have high expectations for its integration, believing that it will enhance the efficiency and quality of government operations.

**3. Digitalization framework in Romania**

Most authors suggest that the various "stages" of digitalization or its "maturity grade" can be seen in the manner in which electronic public services are delivered via the Web having at one end the static display of information (as mentioned in the previous section), and at the other one their delivery in a fully integrated and continuous form (Pardo, 2000; Baltac, 2011; Vrabie, 2009). The characteristics of the latter seem to be focused on a continuous, full, and without difficulty, provision of information within the administrative Web space.

Romania is included on the list of decentralized states, fact which implies that municipalities enjoy a high degree of autonomy in relation to the central public administration (Iancu, 2013), also including here the electronic services' design and management (Baltac, 2011; Vrabie, 2009). At a central level, the digitalization initiatives are coordinated by the Ministry of Research, Innovation and Digitalization, as well as (or in collaboration with) the Ministry of Development, Public Works and Administration. In contrast to other European countries l(i.e., Estonia and Denmark), Romania does not have a position of CIO (Chief Information Officer) both at the central and local levels, which can coordinate digitalization efforts targeting the achievement of inter and intra-institutional collaboration. The absence of a CIO in Romania may lead to a lack of coherent strategy, inefficient use of resources, and slower progress in digitalization efforts (United Nation, 2020; Ojo, et al., 2007; Obi, 2007).

**4. Research methodology**

Romania is the twelfth-largest country in Europe – being located in the Center-East of the continent (Figure 1), and is also the sixth most populous in European Union. The country is divided into 41 counties, each of them having a different number of municipalities (103 in total), cities (217) and communes (2,862).

Municipalities are the most rich administrative units – this being the reason for this article to only focus on them, and they have the widest range of self-government tasks among all. They are responsible for delivering the services that their local community needs. Based on the above, municipalities are directly influencing the digitalization level.

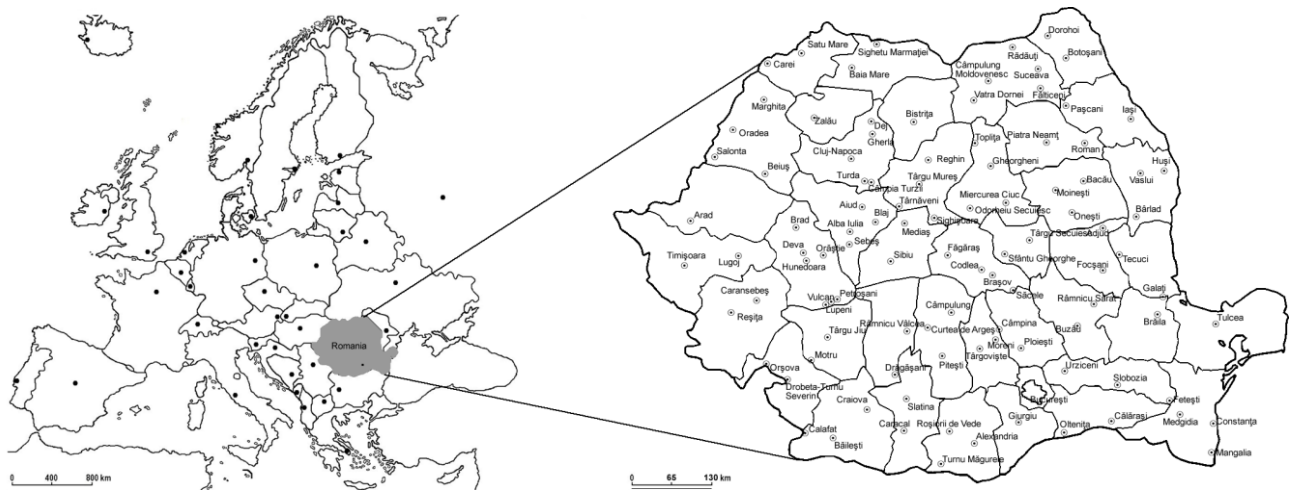

**Figure 1.** (left) Location of Romania in Europe; (right) Counties and municipalities of Romania. Source: author

It is considered, by practice (Feeney & Brown, 2017; Khudeira, n.d.; Vrabie, 2009), that public authorities are promoting the values of their actions firstly by posting them (as an initiative or a running project), on their Web site (Paulin, 2018; Svara, 2021). As early as 1999 Jon M. Kleinberg from Cornell University studied the network structure of a hyperlinked environment and develop a set of algorithmic tools for extracting information from the link structures of such environments (Kleinberg, 1999).

However, the goal of this article is not solely to advocate for personalized digital services as a mandatory step forward. Rather, we aim to explore these services as a "model" for the spread of web technologies, with a specific focus on the most commonly offered services on Romanian city halls' web portals. To achieve a comprehensive understanding, we have analyzed the prevalence of internet-based electronic services in all 103 municipalities in Romania.

The study commenced in 2010, with assessments conducted biennially through to 2014, and was subsequently expanded to encompass the entirety of Romanian cities (103 + 217 in total) by 2016, culminating in a dataset that comprehensively spans the digital maturity of local governance (Vrabie, 2011; Vrabie, 2014; Smart-EDU Hub, 2023). The strategic decision to analyze data from the years 2014, 2019, and 2023 for the present article is predicated upon the observation of modest biennial changes, thus adopting a more extended interval approach to elucidate more significant developmental shifts. Moreover, this temporal framing is deliberately chosen to encapsulate the transformative effects of the COVID-19 pandemic. With the pandemic's onset at the close of 2019 and its continuation into 2023, the analysis is poised to underscore the consequential changes precipitated by this global event, expected to have acted as a potent accelerant for digitalization, compelling municipalities to expedite the adoption of digital services and online engagement in response to the exigencies of remote interaction and public service continuity.

To gather the data for this study, the authors utilized the ParseHub API (ParseHub, 20) to collect information from all 103 official municipal websites simultaneously. To ensure that no data was lost due to websites that were not running during the initial query, the operation was repeated one week later. This approach allowed for a more comprehensive and accurate collection of data.

The study organized the queries into five distinct categories (named from here onwards, classes), which are crucial in evaluating the municipal websites. The first class addressed Transparency, with a focus on how the municipality adheres to regulations concerning this aspect. The second concerned E-Documents and examined various aspects of this topic. The third one, Communication, aimed to extract essential information regarding how the municipality manages the use of new media while the fourth class aimed to provide data on the availability of Useful information online, such as city maps, live cameras, and newsletters. The final class evaluated the ease of browsing the municipality's website as well as other aspects regarding how user friendly is the website. The responses to the queries were then rated on a fixed-point scale, as presented in Table 1.

To conduct the analysis, a total of 48 instruments were used to investigate specific websites. However, for the present article, only 23 of them were considered, based on their relevance to the research question. These 23 instruments were then grouped into five distinct classes – as mentioned above, each with a different number of indicators as presented in Table 1.

**Table 1. Classes of analysis as used for the study**

| C1 – Transparency | C2 – E-Documents | C3 – Communication | C4 – Useful Content | C5 – Generalities |
|---|---|---|---|---|
| C11. Employees declaration of wealth sources | C21. On-line forms and/or off-line (.pdf, .doc, .xls) | C31. Mayor cabinet direct contact line (by email, tel. or WhatsApp number) | C41. City map on Google map platform (updated and maintained by the municipality) | C51. Pleasant design of the city official Web site* |
| C12. Organizational chart | C22. Tracking of submitted application | C32. On-line suggestions of improvement | C42. In site search by keywords | C52. Easy browsing inside the Web site** |

| C13. Minutes of the internal / public meetings | C23. On-line petitioning | C33. Social media official presence (Facebook, Instagram, Twitter etc.) | C43. Multiple language selector | C53. Contact information regarding private companies that provides public services |
|---|---|---|---|---|
| C14. Employees resumes | C24. Public announcements | C34. Sign in / Log in section for citizens | C44. City news section | |
| C15. Budget information | | | C45. List of live cameras Web addresses for citizens to connect | |
| C16. Existence of legislation and city/county decisions | | | C46. Newsletter subscription | |
| **Maximum no. of points per class = 6** | **Maximum no. of points per class = 4** | **Maximum no. of points per class = 4** | **Maximum no. of points per class = 6** | **Maximum no. of points per class = 3** |
| **Maximum no. of points per city = 23** ||||| 

*, ** – While the first four classes are easy to measure ('0' for non-existent information and '1' for existing information) the C5 – Generalities needs some further explanation. Therefore, in Table 2 we present the C51 and C52 subclasses criteria:

**Table 2.** Criteria for C51 and C52 subclasses (R.Anusha, 2014; Alsaeedi, 2020; Mallon, 2014; Bigby, 2018; Craig, n.d.)

| Grade | Description |
|---|---|
| 1 | - the portal's design is very poor, unprofessional, probably executed in house<br>- difficult navigation, the site is built in .html, it has no dynamism, the maximum number of clicks needed to reach the last page of a branch is more than four. |
| 2 | - the design is poor, probably executed in house<br>- difficult navigation, the site is built in .html and it has no dynamism. |
| 3 | - satisfactory design, yet the page is overloaded<br>- difficult navigation, bushy menus, hard to identify exactly where the information can be found / general information about the municipality are displayed in a to „be there" manner. |
| 4 | - pleasant contrasts, airy page / easy navigation, but with bushy menus even if they are executed in advanced programming languages (ASP, PHP etc.)<br>- the information about the municipality are rich and „at sight". |
| 5 | - the Web site is executed in a professional manner, airy<br>- navigation is completely dynamic and intuitive / the information about the municipality are very rich and easy to find. |

Below one can see the formula needed to convert the 1 to 5 scale into points:

$C51 = GC51 \times 0.20$ (1)

$C52 = GC52 \times 0.20$ (2)

Where:
C51, C52 - Values for the named subclasses;
GC51, GC52 - Grades received by each subclass.

The formulas for every class and for the final result used in the present study are as follow:

$C1 \ (TRANSPARENCY) = \sum_{i=1}^{MaxC1} C1(i)$ (3)

$C2 \ (E-DOCUMENTS) = \sum_{i=1}^{MaxC2} C2(i)$ (4)

$C3 \ (COMMUNICATION) = \sum_{i=1}^{MaxC3} C3(i)$ (5)

$$C4 \text{ (USEFUL CONTENT)} = \sum_{i=1}^{MaxC4} C4(i) \qquad (6)$$

$$C5 \text{ (GENERALITIES)} = \sum_{i=1}^{MaxC5} C5(i) \qquad (7)$$

$$Ms = \sum_{i=1}^{MaxC1} C1(i) + \sum_{i=1}^{MaxC2} C2(i) + \sum_{i=1}^{MaxC4} C4(i) + \sum_{i=1}^{MaxC4} C4(i) + \sum_{i=1}^{MaxC5} C5(i) \qquad (8)$$

Where:
C1, C2, C3, C4, C5 - Classes of analysis as in Table 1;
C1(i), C2(i), C3(i), C4(i), C5(i) - Indicators used for investigating the Web site; To build up the result for C5 formula (1) and (2) was used;
MaxC(1-5) - Maximum number of points per class as in Table 1;
Ms - Municipality score – the final score obtained by the Web site.

In order to have a clear perspective on the actual stage of Romanian municipalities official Web sites, we converted the absolute scores received by each class of analysis into a relative 1 to 5 scale (Likert scale, were 1 is showing the lowest score and 5 the highest) using the Excel CEILING function over the following mathematical formula:

$$Relative\ value\ of\ each\ class = \frac{Cji - \min Cj}{\frac{\max Cj - \min Cj}{5}} \qquad (9)$$

Were:
j - Takes value from 1 to 5 according to each class of analysis;
i - Takes value from 1 to 103 according to each municipality.

Below, one can see the Excel formulas used for conversion of absolute values into relative ones on a 1 to 5 scale:

```
IF(C1="","",MIN(MAX(CEILING((C1-MIN(TSoS@C1))/((MAX(TSoS@C1)-MIN(TSoS@C1))/5),1),1),5))  (10)

IF(C2="","",MIN(MAX(CEILING((C2-MIN(TSoS@C2))/((MAX(TSoS@C2)-MIN(TSoS@C2))/5),1),1),5))  (11)

IF(C3="","",MIN(MAX(CEILING((C3-MIN(TSoS@C3))/((MAX(TSoS@C3)-MIN(TSoS@C3))/5),1),1),5))  (12)

IF(C4="","",MIN(MAX(CEILING((C4-MIN(TSoS@C4))/((MAX(TSoS@C4)-MIN(TSoS@C4))/5),1),1),5))  (13)

IF(C5="","",MIN(MAX(CEILING((C5-MIN(TSoS@C5))/((MAX(TSoS@C5)-MIN(TSoS@C5))/5),1),1),5))  (14)

IF(Ms="","",MIN(MAX(CEILING((Ms-MIN(TSoS@Ms))/((MAX(TSoS@Ms)-MIN(TSoS@Ms))/5),1),1),5))  (15)
```

Were:
Ci, Ms - The value obtained by using formulas 3 to 8;
TSoS@Ci - The Total Set of Scores obtained at class Ci where i takes values from 1 to 5 according to the class no.;
TSoS@Ms - The Total Set of Scores obtained by the municipality (Ms).

## 5. Results

Several cities have demonstrated superior performance in one category while failing to achieve high scores in other areas. It is important to note that we have refrained from explicitly naming any city in this study. The rationale behind this decision is to avoid any potential misuse of the article for political purposes. As authors, we intend to maintain a neutral stance and distance ourselves from any political debates that may arise subsequent to publication.

### 5.1. General results

All of the Romanian municipalities do have an active Web site on the Internet. However, what is notable is that a high percentage (95.15%) of these municipalities are also present on social media platforms such as Facebook, Twitter, Instagram, Tumblr, Flickr etc. This suggests that Romanian municipalities are actively engaging with their citizens on social media, which has replaced traditional bulletin boards and forums (Cassel, 2016; Driscoll, 2016; Holt, 2020) as a means for citizens to voice their opinions regarding actions and activities that the city hall is conducting (Tapscott, 2008).

Analyzing Table 3, which outlines the evolution of electronic public services available on Romanian municipalities' websites from 2014 to 2023, one can see the difference, particularly when considering the modest changes between 2014 and 2019, and the substantial advancements by 2023, likely influenced by the COVID-19 pandemic.

**Table 3. Electronic public services available on Romanian cites Web sites**

| Electronic public services | 2014 | | 2019 | | 2023 | |
|---|---|---|---|---|---|---|
| | No. of municipalities | % | No. of municipalities | % | No. of municipalities | % |
| Active Web site for the city hall | 102 | 99.02% | 103 | 100.00% | 103 | 100.00% |
| Official Social Media profile of the city hall | 9 | 08.73% | 15 | 14.56% | 98 | 95.15% |
| Sign in / Log in section for citizens | 23 | 22.33% | 29 | 28.16% | 68 | 66.02% |
| The existence of electronic forms on the Website | 32 | 31.06% | 37 | 35.92% | 91 | 88.35% |
| Online/mobile tracking of submitted applications | 27 | 26.21% | 33 | 32.04% | 76 | 73.79% |
| Online/mobile petitions | 43 | 41.74% | 48 | 46.60% | 88 | 85.44% |
| The citizen's possibility to subscribe to a newsletter | na | na | 16 | 15.53% | 83 | 80.58% |

Data source: authors own compiled data.

A remarkable surge in digital services is observed post-2019, likely catalyzed by the pandemic-driven shift to digital platforms. The percentage of municipalities with an official social media profile soared to 95.15%, and those offering a 'Sign in / Log in' section almost tripled to 66.02%. This suggests that the pandemic played a crucial role in accelerating digital adoption.

Similarly, analyzing Table 4 shows the quality scores of Romanian municipalities' websites from 2014 to 2023, one can see the modest growth between 2014 and 2019 but a substantial improvement post-pandemic (2019-2023) with the number of municipalities rated as 'Very good' jumping from 9.71% to 53.40%. This substantial increase suggests that the pandemic played a pivotal role in accelerating the digital transformation of municipal websites.

**Table 4. Aggregated view over Romanian cites Web sites (relative scores)**

| Grade* | 2014 | | 2019 | | 2023 | |
|---|---|---|---|---|---|---|
| | No. of municipalities | % | No. of municipalities | % | No. of municipalities | % |
| Very good (overall relative score equal to 5) | 10 | 9.71% | 10 | 9.71% | 55 | 53.40% |
| Good (overall relative score equal to 4) | 33 | 32.04% | 37 | 35.92% | 25 | 24.27% |
| Satisfactory (overall relative score equal to 3) | 46 | 44.66% | 52 | 50.49% | 15 | 14.56% |
| Poor (overall relative score equal to 2) | 12 | 11.65% | 3 | 2.91% | 6 | 5.83% |
| Very poor (overall relative score equal to 1) | 2 | 1.94% | 1 | 0.97% | 2 | 1.94% |

* calculated by Formula 15.

In 2023, approximately 90% of municipalities (combining 'Very good' and 'Good' categories) have websites that are well-designed and rich in information, indicating a strong commitment to digital accessibility and public service provision. The achievement of a perfect score (23 points) by one municipality in 2023 is particularly noteworthy. This demonstrates an exceptional commitment to providing a high-quality, user-friendly, and informative digital platform for citizens.

While the period between 2014 and 2019 saw gradual digitalization in Romanian municipalities, the pandemic appears to have significantly accelerated this process. By 2023, there was a marked improvement in the range and sophistication of electronic public services, reflecting a rapid adaptation to the demands of a digital-first era. This shift not only indicates a response to immediate challenges posed by the pandemic but also suggests a more profound and lasting transformation towards digital governance.

**5.2. Results on classes**

Upon examination of Table 5, which delineates the relative scores achieved by Romanian municipalities across various evaluative categories, one notes the incremental improvements from 2014 to 2019, in contrast to the marked progress observed between 2019 and 2023—a period of advancement that can be reasonably attributed to the exigencies imposed by the COVID-19 pandemic. From this data, several important insights emerge:
- Significant Improvement in Transparency: There is a notable increase in the number of municipalities achieving the highest score in 'Transparency' from 2014 to 2023 (from 42.72% to 85.44%). This nearly doubling in percentage indicates a strong commitment to transparency, possibly driven by legislative changes and heightened public expectation for open governance.
- Growth in E-Documents and Useful Content: The 'E-Documents' and 'Useful Content' classes show substantial improvements (from 18.45% to 69.90% and 20.39% to 62.14%, respectively). These increases suggest an enhanced focus on digitalization and the provision of relevant, useful information on municipal websites.
- Communication as a Developing Area: The 'Communication' class shows a moderate increase (from 15.53% to 61.17%). While the improvement is significant, it highlights ongoing efforts and potential areas for further enhancement in how municipalities engage with citizens digitally.
- Generalities Indicating Room for Growth: The 'Generalities' category shows the least improvement (from 9.71% to 27.18%). This suggests that while there is some advancement, there remains considerable scope for municipalities to develop more comprehensive, user-friendly, and interactive online presences.

Overall, the trajectory of scores over the years points to a concerted effort by Romanian municipalities to enhance their digital presence and service delivery. The marked improvements across all categories, particularly post-2019, suggest once again that the pressures and opportunities presented by the COVID-19 pandemic may have served as a catalyst for accelerated digitalization and improvement in public sector services.

**Table 5.** Relative scores obtained by the cities of Romania on each class of analysis

| Class of analysis Score | Year | Transparency | E-Documents | Communication | Useful content | Generalities |
|---|---|---|---|---|---|---|
| 5 | 2023 | 88 (85.44%) | 72 (69.90%) | 63 (61.17%) | 64 (62.14%) | 55 (27.18%) |
|   | 2019 | 46 (44.66%) | 22 (21.36%) | 18 (17.48%) | 23 (22.33%) | 11 (10.68%) |
|   | 2014 | 44 (42.72%) | 19 (18.45%) | 16 (15.53%) | 21 (20.39%) | 10 (9.71%) |
| 4 | 2023 | 7 (6.80%) | 16 (15.53%) | 34 (33.01%) | 21 (20.39%) | 25 (41.75%) |
|   | 2019 | 36 (34.95%) | 16 (15.53%) | 56 (54.37%) | 24 (24.27%) | 35 (33.98%) |
|   | 2014 | 35 (33.98%) | 13 (12.62%) | 54 (42.43%) | 25 (24.27%) | 33 (32.04%) |
| 3 | 2023 | 0 (0.00%) | 0 (0.00%) | 0 (0.00%) | 14 (13.59%) | 15 (22.33%) |
|   | 2019 | 15 (14.56%) | 26 (25.24%) | 0 (0.00%) | 27 (26.21%) | 24 (23.30%) |
|   | 2014 | 17 (16.50%) | 27 (26.21%) | 0 (0%) | 31 (30.10%) | 26 (25.24%) |
| 2 | 2023 | 7 (6.80%) | 6 (5.83%) | 5 (4.85%) | 1 (0.97%) | 6 (7.77%) |
|   | 2019 | 4 (3.88%) | 21 (20.39%) | 26 (25.24%) | 19 (18.45%) | 26 (24.24%) |
|   | 2014 | 5 (4.85%) | 23 (22.33%) | 28 (27.18%) | 17 (16.50%) | 26 (25.24%) |
| 1 | 2023 | 1 (0.97%) | 9 (8.74%) | 1 (0.97%) | 3 (2.91%) | 2 (0.97%) |
|   | 2019 | 2 (1.94%)) | 18 (17.48%) | 3 (2.91%) | 9 (8.74%) | 7 (6.80%) |
|   | 2014 | 2 (1.94%) | 21 (20.39%) | 5 (4.85%) | 9 (8.74%) | 8 (7.77%) |

Data source: authors own compiled data from 2014, 2019 and present study.

### 5.3. Auxiliary results

Previous studies (Moon & Peter deLeon, 2001; Moon, 2002; Juliet Musso, et al., 2000) have suggested a positive relationship between a city's population size and the e-governance capabilities of its local public administration. With the extensive data gathered in this research, we aimed to confirm this positive correlation for Romania as well. As shown in Figures 2. and Figure 3., the correlation coefficient is 0.48 (as in the 2023 study), which is statistically significant, indicating a moderately positive relationship between the two data sets (municipality population and grades obtained from the analysis).

It is important to note that Bucharest, the capital of Romania, was excluded from this calculation due to the risk of skewing the results, as it accounts for more than 10% of the country's population.

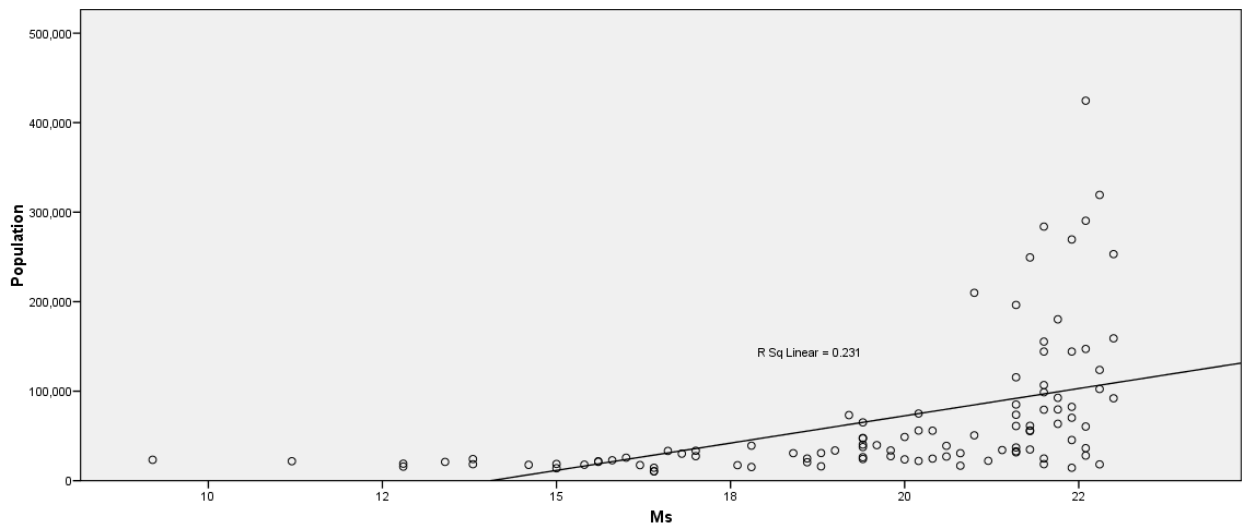

**Figure 2.** Scatter Plot Analysis: Population in each municipality (1). The linear regression line, with an $R^2$ value of 0.231, suggests a weak positive correlation between the municipality identifier and population size. Data source: authors own compiled data.

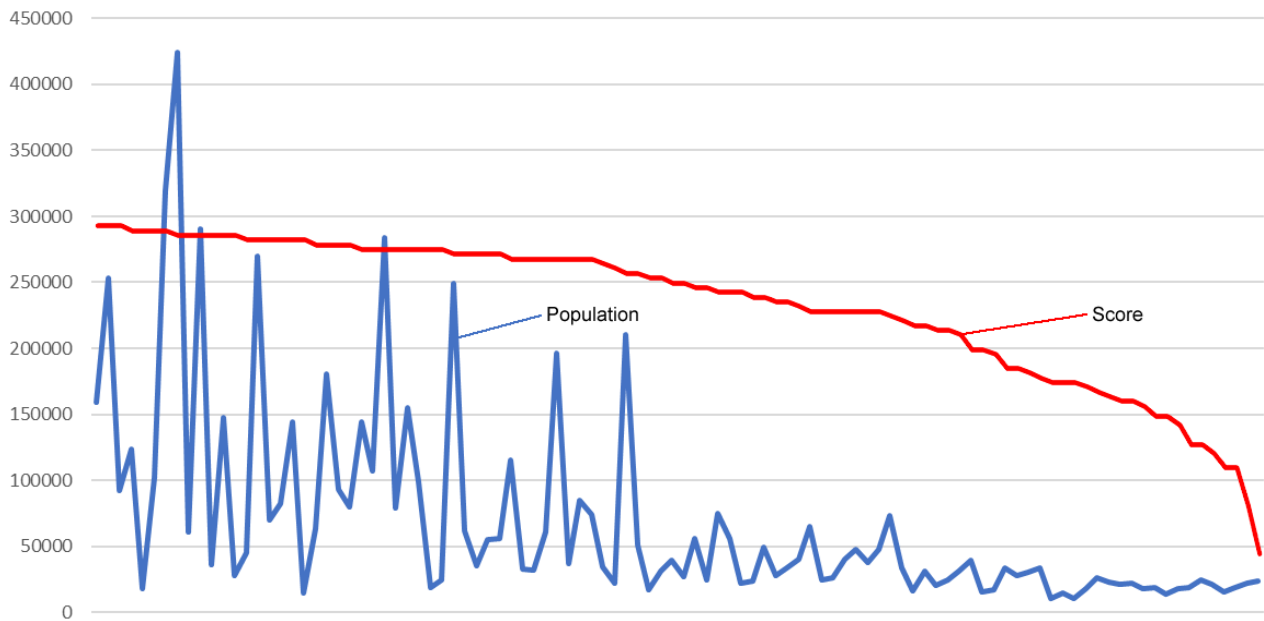

**Figure 3.** The correlation between the population and the score obtained by municipalities (2). Data source: authors own compiled data.

## 6. Study limitation

While the study presented above offers valuable insights into the relationship between city population size and e-governance capabilities, there are several limitations that should be considered when interpreting the results:

- Nature of the data: The study's national focus makes replication impossible. Although the questionnaire was applied to all Romanian municipalities, which allowed for control of potentially intervening and confounding factors, reducing the risk of endogeneity, this limitation still remains.
- Aging data: The data used in the study may no longer accurately reflect the current state of e-governance capabilities in Romanian municipalities, as there is a high likelihood that some municipalities have improved their online presence since the data was collected. The rapid growth of the digital landscape during and after the Covid era in Romania and other countries may have further exacerbated this issue.
- Inability to make international comparisons: Comparing the results of this study with those from other countries in Europe or worldwide may not be accurate due to differences in economies, policies, approaches, and general conditions that influence the pace of digitalization.

- Correlation does not imply causation: Although a positive correlation was found between population size and e-governance capacity, this does not necessarily mean that larger population size directly leads to better e-governance. There may be other factors, such as economic development or technological infrastructure, that contribute to both population size and e-governance capacity.
- Limited indicators: The study relied on specific indicators to assess e-governance capacity, which may not capture all aspects of local public administration's capabilities. The use of additional or alternative indicators could potentially yield different results.
- Potential for measurement error: The process of collecting data for some indicators (i.e., C51, C52 – as in Table 1) and assigning scores to them, may be subject to human error or biases being made manually. Ensuring the reliability and validity of the data collection and scoring process is crucial for the accuracy of the findings.
- Primarily concentration on assessing the level of online engagement: This focus may not capture the full spectrum of interactions and communication channels that exist between citizens and government bodies. Moreover, the research may not account for potential barriers to online engagement, such as digital literacy, internet access, or the effectiveness of the digital tools provided by the institutions. As a result, the findings might not fully reflect the overall engagement and interaction between citizens and their local governments.

By addressing these limitations in future research, a more comprehensive understanding of the relationship between city population size and e-governance capabilities can be achieved.

## 7. Findings

Smart cities optimism and e-government 3.0 represent the latest trends in utilizing advanced technologies to enhance urban living, governance, and public services. From the perspective of municipalities' web pages, these concepts can be translated into the adoption of various digital tools, solutions, and strategies that improve the communication between citizens and local governments and streamline the provision of public services.

- Enhanced User Experience: Municipalities' web pages should be designed with a user-centric approach, providing easy navigation, engaging content, and an intuitive interface that encourages citizens to participate in governance processes and access services online.
- Integrated Services: E-government 3.0 envisions seamless integration of public services across different departments and agencies, allowing citizens to access services more efficiently. Municipalities' web pages can act as one-stop portals, connecting various departments and providing access to multiple services from a single platform.
- Data-driven Decision-making: Smart cities rely on data for better decision-making and optimized resource allocation. Municipalities can integrate data analytics tools and dashboards that offer insights into urban trends, patterns, and challenges, enabling local governments to make informed decisions and develop targeted policies.
- Open Data and Transparency: E-government 3.0 emphasizes transparency and openness in governance. Municipalities' web pages can act as platforms for sharing open data, budget information, and legislative updates, promoting accountability and encouraging citizen participation in the decision-making process.
- Digital Participation and Collaboration: Municipalities' web pages can facilitate digital participation by offering electronic referendums, and public consultations. This enables citizens to have a direct impact on the governance process and fosters a sense of ownership and responsibility among the community.

In conclusion, the optimism surrounding smart cities and e-government 3.0 can be reflected in municipalities' web pages by adopting new technologies, user-centric design, and innovative solutions that enhance citizen engagement, promote transparency, and streamline public services.

## 8. Discussions and conclusions

The article presents the state of digitalization in Romania as observed through the websites of its municipalities. If we are to zoom in in time, or to see the e-government evolution as a pyramidal structure, it can be posited that the ICT sector is steadily advancing toward maturity. This progression paves the way for the imminent e-gov 3.0, characterized by the expanded integration of artificial intelligence (AI) tools within government operations. The incorporation of AI is anticipated to enhance and streamline various processes, ultimately contributing to the overall growth and development of the e-government landscape.

Similar research conducted in other countries includes a wider range of indicators, such as online payments (Holzer & Aroon P. Manoharan, 2016; Homburg & Andres Dijkshoorn, 2011; Gonzalez, et al., 2007), citizen participation in governance through electronic voting or electronic referendums (Tapscott, 2008; Chaieb, et al., 2018) and online surveys designed to gather public opinions on potential actions by the city hall (Holzer & Aroon P. Manoharan, 2016; Homburg & Andres Dijkshoorn, 2011).

A significant correlation is observed between the outcomes derived from each analytical category and the ultimate result (Table 6). Nevertheless, the study endeavored to ascertain which facet exerted the most substantial influence on the final outcomes. The 'Generalities' category exhibited the lowest correlation at 0.295, which can be primarily attributed to the fact that websites are often developed in-house rather than by professionals, resulting in the omission of numerous crucial features.

**Table 6.** Correlation between the Final result (2023) and the classes of analysis

| Class analyzed | Pearson Correlation with the final results (Cj & Ms) |
|---|---|
| Transparency | 0.740 |
| E-Documents | 0.845 |
| Communication | 0.543 |
| Useful content | 0.842 |
| Generalities | 0.295 |

Definitely, if compared with previous years and studies, the COVID-19 pandemic has acted as a catalyst for the development and adoption of Information and Communication Technology (ICT) within various sectors, including e-government. As governments worldwide grappled with the challenges posed by the pandemic, the demand for efficient and contactless service delivery soared. The necessity for social distancing measures and remote work arrangements accelerated the integration of digital tools and platforms, expediting the transition towards a more sophisticated e-government infrastructure. This unexpected global crisis underscored the importance of leveraging ICT to ensure the continuity of essential public services, while simultaneously safeguarding the well-being of citizens and government personnel. Consequently, the pandemic has served as a powerful impetus for the expansion and maturation of ICT, emphasizing its crucial role in the resilience and adaptability of governments in times of crisis as shown within the present article.